# Feedback Scheduling of Priority-Driven Control Networks

Feng Xia[a,b], Youxian Sun[a], Yu-Chu Tian[b]

[a]*State Key Laboratory of Industrial Control Technology, Zhejiang University*
*Hangzhou 310027, China*
[b]*Faculty of Information Technology, Queensland University of Technology*
*GPO Box 2434, Brisbane QLD 4001, Australia*
Emails: f.xia@ieee.org; yxsun@iipc.zju.edu.cn; y.tian@qut.edu.au

**Abstract**
With traditional open-loop scheduling of network resources, the quality-of-control (QoC) of networked control systems (NCSs) may degrade significantly in the presence of limited bandwidth and variable workload. The goal of this work is to maximize the overall QoC of NCSs through dynamically allocating available network bandwidth. Based on codesign of control and scheduling, an integrated feedback scheduler is developed to enable flexible QoC management in dynamic environments. It encompasses a cascaded feedback scheduling module for sampling period adjustment and a direct feedback scheduling module for priority modification. The inherent characteristics of priority-driven control networks make it feasible to implement the proposed feedback scheduler in real-world systems. Extensive simulations show that the proposed approach leads to significant QoC improvement over the traditional open-loop scheduling scheme under both underloaded and overloaded network conditions.

**Keywords:** Networked control systems, control networks, feedback scheduling, dynamic bandwidth allocation

## 1. Introduction

Networked control systems (NCSs) are becoming increasingly important in modern control engineering and applications [6,16]. An NCS uses a control network [13,15,19,20] to interconnect geographically distributed nodes such as sensors, controllers, and actuators. Compared to traditional control systems with point-to-point interconnections, NCSs are advantageous in terms of simple and fast implementation, ease of system maintenance, and increased system flexibility and dependability. However, the use of communication networks in control applications complicates the analysis and design of the control systems. The resulting performance of the NCS systems depends heavily on temporal network attributes such as network-induced delay, packet loss, and jitter [9,21-23], which are closely related to the runtime availability of network bandwidth.

From the communication perspective, to satisfy the requirements of control applications, it is often necessary for control networks to provide deterministic real-time communications. This poses a technical limitation on the maximum possible transmission rate that the control networks can offer. For example, the Controller Area Network (CAN) bus has the maximum transmission rate of 1Mb/s [13,20,28]. Control networks with much higher data rate are available now, but such networks are often very expensive and thus are not an economically viable solution for most industrial applications. Therefore, it is common in real-world applications that the bandwidth of control networks is limited, making the available communication resource a performance bottleneck [6,27].

This bottleneck is further accentuated by the fact that NCSs are becoming more and more complex and the number of control loops attached to a shared control network continues to grow [16]. Moreover, NCSs often have to operate in dynamic environments where the workload varies due to runtime system reconfiguration or update, which are often required to make the system flexible enough to meet stringent requirements of the changing market [14].

A natural result of limited network bandwidth and variable workload is the uncertainty in available communication resources [27,30]. In terms of temporal attributes, it results in unpredictable communication delay, packet loss, and jitter, which will deteriorate the QoC performance of NCSs, and even jeopardize system stability in extreme circumstances. Consequently, the overall performance of a multi-loop NCS depends not only on the design of the control algorithms, but also on the allocation and scheduling of the shared network bandwidth.

This paper is devoted to maximizing the overall QoC of NCSs closed over priority-driven control networks by means of flexible management of network resources. Following the emerging methodology of feedback scheduling [2,27,31], an integrated feedback scheduling (IFS) scheme is proposed that enables flexible QoC management in NCSs subject to bandwidth limitation and workload fluctuation. Unlike most existing NCS solutions where controller algorithms or network MAC protocols are the focus, our approach concentrates on codesign of feedback control and network scheduling. To make the best possible use of available network resources, the IFS scheme adapts simultaneously the sampling periods and the priorities of control loops at runtime. A cascaded feedback scheduling algorithm based on deadline miss ratio control is developed for adjusting sampling periods, and a direct feedback scheduling algorithm is developed for priority modification. In contrast to traditional open-loop scheduling methods for NCSs, our approach features closed-loop scheduling of network resources, which exploits feedback control technology.

The rest of this paper is organized as follows. Related work is briefly reviewed in Section 2. Section 3 describes the system we consider, and introduces the problem of network scheduling, from the viewpoint of integrated control and scheduling. In Section 4, the framework of the IFS scheme is presented, and the algorithms are given for adjusting sampling periods and modifying priorities, respectively. Section 5 evaluates the performance of the proposed approach through extensive simulations. Finally, Section 6 concludes the paper.

## 2. Related Work

Recent overviews on NCSs can be found in [6] and related articles in the same issue. Roughly speaking, the majority of existing work on NCSs can be divided into three main categories: 1) control theoretic approaches, 2) network design based approaches, and 3) control and network codesign. Intuitively, the focus of control theoretic approaches is on controller design, i.e., to design control algorithms that are tolerant to network-induced delay, packet loss, and jitter. The approaches based on network design mainly deal with how to improve network quality-of-service (QoS) such that the control system performance is guaranteed. In this work we are interested in the third category, i.e., the codesign of control and network scheduling.

Network scheduling algorithms that exploit the communication principles of priority-driven control networks have been presented, e.g. [4,8,26,28,29,32]. With focus on the problem of how to schedule messages from different nodes, however, none of these algorithms were developed to attack the variations in available communication resources in NCSs operating in dynamic environments. With these algorithms, deadline misses under overload conditions cannot be avoided, and the waste of resource caused by light workload cannot be reduced, which may potentially yield worse-than-possible QoC. Heuristic methods for allocation of shared bandwidth among multiple control loops have also been developed in e.g. [7,10,17], but they are all open-loop solutions and therefore are not suitable for dealing with dynamic changes in available network resources.

Recently, effort has been made on closed-loop network scheduling that features dynamic bandwidth allocation via sampling period adaptation, e.g. [1,3,11,24,25]. Most of them adjust the sampling periods of control loops to maximize the overall QoC under the constraint on a pre-set level of network utilization. In dynamic environments, however, some resources will be wasted because pessimistic utilization setpoints must always be chosen so as not to violate the system schedulability. An emerging technology for dynamic resource allocation is feedback scheduling, which typically employs feedback control theory and technology to increase flexibility and master uncertainty in various computing systems [2,5,31]. While most of the work on feedback scheduling is dedicated to codesign of control and CPU scheduling, the main concern of this paper is control network scheduling.

In contrast to all of the above-mentioned work, this paper is concerned with integrated feedback scheduling that features simultaneous adaptation of the sampling periods and the priorities of multiple networked control loops, with the goal of improving the overall QoC of the whole system. With period adjustment that takes advantage of feedback control theory and technology, the dynamic changes in available network resource will be addressed. The network bandwidth will be fully utilized even if the original workload is light. Also, overload conditions will be handled with graceful QoC degradation. The dynamic assignment of priorities will further optimize the distribution of available network resources. A preliminary framework of integrated feedback scheduling has been explored in our previous work [30], and will be substantially extended in this paper.

## 3. Problem Statement

### 3.1. System Model

Consider an NCS where $N$ independent control loops share a control network. In this network a priority-driven MAC protocol is employed. Commonplace examples of this type are CAN and DeviceNet. From the principle of these communication protocols, every node device that has packets to send, i.e., the so-called communication entity will be assigned a unique priority level. A node with a packet to send waits until the network is idle and then commences to transmit. In the case of network access collisions, the system will decide which packet will be transmitted according to their priority levels. As usual the packet with the highest priority will be transmitted successfully. The proposed approach is also applicable to NCSs containing interfering traffic, though this is not elaborated on in this paper for simple description.

In the NCS, each control loop is composed of a sensor, a controller, and an actuator, in addition to a controlled process. Assume that the controller and the actuator are connected directly, implying that only the sensor in the control loop needs to use the control network to deliver sample data to the controller. With this system architecture, the priority of a sensor node can be viewed as the priority of the corresponding control loop. It is also assumed that all controllers and actuators are event triggered, while sensors are time triggered.

In a single control loop, the sensor collects a sample of the output of the physical process at the beginning of every sampling period, and then sends it to the controller via control network after successfully accessing. Once receiving the sample data, the controller starts to execute the corresponding control algorithm immediately, producing control signal, and then outputs it into the actuator. Finally the actuator acts on the controlled physical process according to the control input. Assume that in this process the processing delay of the sensor and the actuator and the execution time of the control algorithm are relatively small and thus are neglected. In this context the control delay is approximately equal to the communication delay including mainly waiting delay and transmission delay.

No specific compensation methods for delay, packet loss, or jitter are used in the control loops. However, controller parameters will be updated accordingly when sampling periods are changed. Without loss of generality, it is also assumed that:
- The control network is ideal in that data communications via the control network are error-free.
- Sample data is delivered in the form of single packets, which means that every sample will be treated as one data packet while being transmitted over the network.
- The sensor and the actuator in one loop hold a precisely synchronized clock. Since only the sensors are time triggered in the system, this assumption is not a necessity. The purpose of making it here is simply for exact calculation of the deadline miss ratio.

### 3.2. Control Network Scheduling

Since the control network is shared by multiple communication entities, it is necessary and important to allocate network bandwidth properly, particularly when the transmission rate is limited. In the following the problem of control network scheduling is formulated from the perspective of real-time scheduling.

Generally speaking, the scheduling of networks is similar to the problem of real-time CPU scheduling [3,21]. Both of them consider how to distribute shared resources among a set of concurrent tasks, which are often subject to real-time constraints, either hard or soft. The shared resource to be distributed is no longer CPU time, but network bandwidth in NCSs. The tasks to execute in this context will not be software programs as in CPU scheduling, but messages in the form of data packets over the network. Accordingly, the meaning of task execution changes from running programs to transmitting data packets. Based on these observations, it is found that existing real-time scheduling theory and methods may be applied to control network scheduling by re-defining relevant task attributes. We define the following timing attributes for sample data messages in NCSs:
- *Period $h_i$*: the period for generating a new sample data packet in the *i*-th control loop, equal to the relevant sampling period;
- *Relative deadline $d_i$*: equal to $h_i$;
- *Execution time $c_i$*: the transmission delay of a data packet, excluding waiting delay;
- *Priority $p_i$*: the sensor's priority in the *i*-th loop;
- *Network utilization $U_i$*: $U_i = c_i / h_i$.

Traditional real-time scheduling algorithms such as Rate Monotonic (RM) and Earliest Deadline First (EDF) are theoretically applicable to control network scheduling based on this mapping. Due to practical difficulty in implementation, however, most often fixed-priority scheduling methods are used in real-world NCSs. The most notable difference between CPU scheduling and network scheduling is that in general the execution of programs on CPU is preemptive, whereas the transmission of data over network is not. Once a data transmission starts, it will continue until it is completed, and will never be suspended because of new transmission requests with any priorities.

From the work by Sha *et al.* [18], we can derive the following theorem describing a sufficient condition for NCS schedulability analysis.

**Theorem 1** For an NCS with *N* independent control loops, where an ideal priority-driven control network is used, the system is schedulable with RM if (1) is satisfied.

$$\frac{c_1}{h_1} + \frac{c_2}{h_2} + \cdots + \frac{c_i}{h_i} + \frac{b_i}{h_i} \leq i(2^{1/i} - 1), \quad \forall i = 1, ..., N \tag{1}$$

where $h_1 \leq h_2 \leq \cdots \leq h_N$, $b_i$ is the worst-case blocking time of task *i*, i.e., $b_i = \max_{n=i+1,...,N} c_n$.

Based on the above schedulability constraint, the network scheduling problem in NCSs can be stated informally as follows: *In the presence of limited bandwidth and variable workload, dynamically allocate available communication resource among multiple control loops so that the overall QoC is maximized while meeting the constraint on system schedulability*. In the next section we will present an integrated feedback scheduling approach to address this problem.

## 4. Integrated Feedback Scheduling

In contrast to traditional open-loop network scheduling methods, a closed-loop dynamic network bandwidth allocation method will be developed below. Flexible QoC management will be facilitated by means of codesign of feedback control and network scheduling. The architecture of the IFS scheme will be first described; then the feedback scheduling algorithms for period adjustment and priority modification will be presented, respectively. Some critical design issues will also be discussed.

### 4.1. Architecture

To enable feedback scheduling, it is necessary to make choice of some related variables first. From the principle of feedback control, *controlled variable* and *manipulated variable* are two most important variables that need to be defined in the feedback scheduling system.

In terms of real-time scheduling, there are generally two options for the controlled variable, *network utilization* and *deadline miss ratio* (or *miss ratio* in short). In network scheduling, however, it is very hard, if not impossible, to determine an exact schedulable utilization upper bound. Consequently, it will be difficult to choose an appropriate network utilization setpoint, especially when the network workload varies with time. On the other hand, feedback scheduling based on deadline miss ratio control does not depend on the knowledge about schedulable network utilization upper bound, which avoids the difficulty with network schedulability analysis. Moreover, it is intuitive that controlling deadline miss ratio at a specific (relatively low) level will certainly maintain the actual network utilization close to the highest possible level, regardless of changes in system workload. Because there is always certain stability margin in practical control systems design, real-world control systems can tolerate packet losses (and also deadline misses) to certain degree [3,9,22]. Therefore, the deadline miss ratio is selected as the output of the feedback scheduling system. With regard to the manipulated variable, since the transmission time of sample data packets cannot be intentionally adjusted, the sampling periods of control loops become a natural choice.

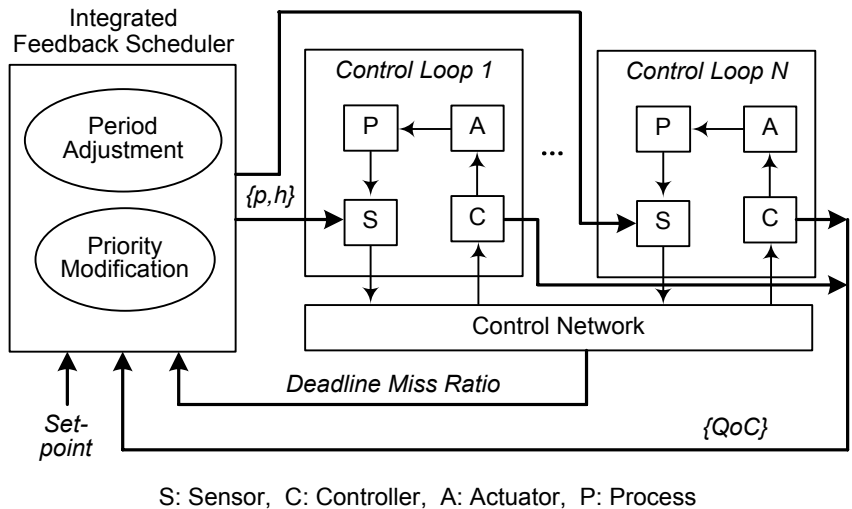

S: Sensor, C: Controller, A: Actuator, P: Process

**Fig. 1.** Architecture of the integrated feedback scheduling

Fig. 1 depicts the architecture of the integrated feedback scheduling. In addition to the original control loops, an outer loop is introduced to implement the feedback scheduling. The feedback scheduler consists of two main components: *period adjustm*ent and *priority modification*. Just as their names imply, the period adjustment module is responsible for dynamic adjustment of sampling periods of the control loops according to network condition and system performance, and the priority modification module re-assigns the sensors' priorities online according to actual performance of the control loops. The inputs to the integrated feedback scheduler include the actual deadline miss ratio and its setpoint, and the measured control performance in terms of certain performance metric. The output variables are the sampling period and the priority of each control loop.

The feedback scheduler is time triggered. At every invocation instant, the scheduler gathers current deadline miss ratio and the control performance of control loops. With respect to the setpoint of the deadline miss ratio, new sampling periods will be produced. Priorities of the control loops will also be modified when necessary.

Intuitively, actual network utilization could be kept around the highest possible level by adapting sampling periods, as long as miss ratio setpoint is not zero. Due to this non-zero miss ratio setpoint, however, deadline misses are unavoidable when the system is in steady states, no matter at what level the workload is. In NCSs that use fixed-priority scheduling methods, most of the deadline misses will happen in control loops whose priorities are relatively low. This may deteriorate the performance of these control loops. This problem will be addressed in the priority modification module.

### 4.2. Period Adjustment

The primary role of the period adjustment module is to adapt the sampling periods of the control loops in a way that the actual deadline miss ratio is maintained at a desired level and the available resources are reasonably distributed. A natural method to construct this module following feedback control theory is to use a specific control algorithm to obtain the sampling periods directly from the difference between actual deadline miss ratio and its setpoint. A method using this idea has been discussed in our previous work [30]. In this paper, we develop a more generalized cascaded feedback scheduling method to determine the sampling periods.

New sampling periods are produced online with two consecutive steps, thus forming a cascaded feedback scheduling algorithm. Firstly, use the classical PID (Proportional-Integral-Derivative) control technique to calculate the total network utilization of all control loops in response to current control error of the deadline miss ratio. Secondly, under this constraint on allowable network utilization, obtain new sampling periods by taking into account actual performance of the control loops.

Let $T_{FS}$ be the invocation interval of the feedback scheduler. Deadline miss ratio $\rho(j)$ is defined as the ratio of the number of sample data packets that miss their deadlines to the total number of sample data packets that are generated in control loops in the time interval $[(j-1)T_{FS}, jT_{FS}]$, where $j$ denotes the invocation instant of feedback scheduler. Let $\rho_r$ be the desired deadline miss ratio. At the $j$-th instant, the total network utilization of control loops $U$ ($0 \leq U \leq 1$) is calculated as:

$$\Delta U(j) = K_P \cdot (ERR(j) - ERR(j-1)) + K_I \cdot ERR(j)$$
$$U(j) = U(j-1) + \Delta U(j) \qquad (2)$$

where $K_P$ and $K_I$ are essential coefficients for the PI control algorithm, $ERR$ is the control error of deadline miss ratio.

The calculation of $ERR$ is worth discussing. In most situations the control error can be computed as $ERR(j) = \rho_r - \rho(j)$ according to its general definition. However, deadline miss ratio is subject to saturation, that is, it could never be a negative. When the network workload is light, the deadline miss ratio will be zero all the time, regardless of changes in workload. On the other hand, a relatively small setpoint for deadline miss ratio is always preferable in order to minimize the impact of deadline misses on QoC. Consequently, in the case of light workload when almost no deadline is missed, the absolute value of the control error will remain small, no matter how low the workload level actually is. This may cause the transient process of the feedback scheduling system to be too slow, thus impairing its dynamic behavior.

To address this problem, a deadzone-based control technique is employed here, and the calculation of the control error is re-formulated accordingly. The feedback scheduling system is considered to be steady if deadline miss ratio falls into the interval $(0, \rho_r]$. Accordingly, if $0 < \rho(j) \leq \rho_r$, then the control error of deadline miss ratio is 0. To improve the dynamic behavior of feedback scheduling, the following equation is used to calculate $ERR(j)$:

$$ERR(j) = \begin{cases} \rho_r & \text{if} \quad \rho(j) = 0 \\ 0 & \text{if} \quad \rho_r \geq \rho(j) > 0 \\ -\rho(j) & \text{if} \quad \rho(j) \geq \rho_r \end{cases} \qquad (3)$$

Once the total utilization $U(j)$ is obtained, it is distributed among multiple control loops in a way that the overall QoC is optimized. Let $J_i$ be the performance index for the $i$-th control loop. The problem of finding the optimal sampling periods can be described as the following optimization problem:

$$\min_{h_1, \ldots, h_N} J = \sum_{i=1}^{N} w_i J_i$$
$$\text{s. t.} \quad \sum_{i=1}^{N} c_i / h_i \leq U(j) \qquad (4)$$

where $w_i$ is the weighting coefficient of loop $i$. It is apparent that the above equation gives a typical optimal feedback scheduling problem [2,12,27,31], of which the basic idea is to minimize the total control cost of the system through adjusting the sampling periods of control loops under the constraint of current total utilization. Recall that the bigger the control cost the worse the QoC.

Using the above formulation, the results of the optimal sampling periods will tightly rely on the forms of the performance indices of the control loops. Generally speaking, $J_i$ could be either stationary (i.e., independent of time) or dynamic (i.e., time-varying). In the time domain, there are commonly three types of options for $J_i$, i.e., infinite-time, finite-time, and instantaneous performance indices. For the sake of simplicity, the absolute instantaneous control error is employed here as the dynamic performance index for QoC, i.e.,

$$J_i(j) = |e_i(j)| \qquad (5)$$

where $e_i$ is the control error within the $i$-th control loop.

With (5), Eq. (6) is used to compute the sampling period of loop $i$.

$$h_i = \frac{c_i}{U_i} = \frac{c_i}{U_{i,\min} + U'_i} = \frac{c_i}{U_{i,\min} + (U - \sum U_{n,\min})\frac{w_i J_i}{\sum w_n J_n}}$$

$$= \frac{c_i \cdot \sum_{n=1}^{N} w_n J_n}{\frac{c_i}{h_{i,\max}} \cdot \sum_{n=1}^{N} w_n J_n + w_i J_i (U - \sum_{n=1}^{N} \frac{c_n}{h_{n,\max}})} \quad (6)$$

where $h_{i,max}$ is the maximum allowable sampling period, and $U_{i,min} = c_i/h_{i,max}$ is the corresponding minimum allowable utilization. In Eq. (6), the indicator $j$ for the feedback scheduler invocation instant is omitted for notation simplicity.

Once the minimum utilization of each control loop has been preserved, Eq. (6) will distribute the remaining fraction of bandwidth resource, i.e., $U(j) - \sum_{n=1}^{N} U_{n,\min}$, among the control loops in proportion to the magnitudes of $w_i J_i$. In this way, the control loops with worse performance will be assigned larger fractions of free network bandwidth, while the sampling periods of the control loops with better performance will be more close to their maximum allowable values. In the extreme, if $J_i = 0$ indicating that the control loop is in a steady state, then it can be deduced from (6) that $h_i = h_{i,max}$, implying that the sampling period of this loop is set to the maximum. With this bandwidth allocation scheme, the largest fraction of free communication resource is distributed to the control loop that needs it most. This benefits optimizing the overall QoC.

Eq. (6) assumes that $\sum w_n J_n \neq 0$. In the case where all control loops are in steady states, i.e., $\sum w_n J_n = 0$, the free bandwidth will be distributed among all control loops evenly. Furthermore, to avoid too big difference between the network utilization of the control loops when $\sum w_n J_n$ is small, which may be caused by the fact that some control loops are steady while the others are not, the following rule is introduced into the calculation of $h_i$:

$$h_i = \frac{c_i}{U_i} = \frac{c_i}{U_{i,\min} + U'_i} = \frac{c_i}{U_{i,\min} + (U - \sum U_{n,\min})/N} \quad \text{if } \sum w_n J_n < \varepsilon \quad (7)$$

where $\varepsilon$ is a user-defined parameter.

### 4.3. Priority Modification

The reasons for modifying the priorities of the control loops online together with sampling period adjustment are explained roughly from the following two aspects:
1) To alleviate the effect of deadline misses. As mentioned previously, the feedback scheduling method based on deadline miss ratio control will unavoidably induce deadline misses in the control loops, and the performance of the control loops with lower priorities will be influenced. To address this problem, the priorities of the sensors are modified dynamically according to feedback about the control performance and assign lower priorities to the control loops with better performance. In this way, the deadline misses will most likely occur in the control loops with the best performance, which benefits reducing the impact of deadline misses on the overall QoC.
2) To alleviate the effect of delay. According to the principle of priority-driven communication protocols, high-priority data packets will be transmitted earlier than the data packets with lower priorities in the case of media access contention. A natural consequence is that the control delay of a high-priority loop is usually shorter than that of a low-priority loop. Longer delays often make control performance worse. Furthermore, for given delays, a control loop with better performance could intuitively be impacted less significantly than a control loop with worse performance. Based on this observation, higher priorities are assigned to the control loops that currently have worse performance so that the impact of delay is alleviated.

Of course, the ultimate goal of priority modification is to improve the overall control performance. The basic rule used to assign priorities is that the worse the current performance of a control loop is, the higher the priority it will be assigned. As a prerequisite for priority re-assignment, an appropriate performance metric $J'_i$ should be determined. Similar to the choice of $J_i$, there are many different feasible forms for $J'_i$. Here we define $J'_i$ as follows based on (5):

$$J'_i(j) = w_i J_i(j) = w_i \cdot |e_i(j)| \qquad (8)$$

Using this performance index, the rule for determining priorities turns to be that the bigger the value of $J'_i$ the higher the relevant control loop's priority. Intuitively, for control loops with the same $J'_i$, the order of their priority levels will not be changed.

Since $J_i$ varies over time, it is possible that the priorities of the control loops might be modified frequently due to even small changes in perturbations in some control loops. To reduce the number of unnecessary switches of priorities caused by small variations of $J'_i$ values and to avoid too many fluctuations, the notion of *priority switch threshold $\delta$* is introduced. With this notion, priority switches are permitted only when the absolute difference between the corresponding $J'_i$ values is no less than $\delta$. Accordingly, the following rules are built for priority modification, where *m* and *n* are indexes of control loops.

**RULE 1** If $\max\{J'_i\} - \min\{J'_i\} < \delta$, then make no change of all priority levels.

**RULE 2** If $J'_m(j) = J'_n(j)$, then maintain the order of current priority levels.

**RULE 3** If $J'_m(j) > J'_n(j)$ and $p_m(j-1) > p_n(j-1)$, assuming there is no any control loop whose $J'_i$ value is between $J'_m(j)$ and $J'_n(j)$ (hereafter the same), then $p_m(j) > p_n(j)$.

**RULE 4** If $J'_m(j) > J'_n(j)$, $p_m(j-1) < p_n(j-1)$ and $J'_m(j) - J'_n(j) \geq \delta$, then $p_m(j) > p_n(j)$.

**RULE 5** If $J'_m(j) > J'_n(j)$, $p_m(j-1) < p_n(j-1)$ and $J'_m(j) - J'_n(j) < \delta$, then $p_m(j) < p_n(j)$ and $p_\ell(j) < p_m(j)$, where $\ell$ is the index of the control loop whose $J'_i$ value is the maximum among all that are small than $J'_n(j)$.

The introduction of RULE 5 is to simplify the process of priority re-assignment in some particular cases where $J'_m(j) > J'_n(j) > J'_\ell(j)$ and $p_m(j-1) < p_n(j-1) < p_\ell(j-1)$. No claim is made that $p_m(j) = p_m(j-1)$ and $p_n(j) = p_n(j-1)$ when $J'_m(j) = J'_n(j)$. Since only the order of their priority levels is maintained, the values of $p_m(j)$ and $p_n(j)$ may consequently change with the relative $J'_i$ value of each control loop.

In terms of feedback scheduling, the above priority modification method can be regarded as some kind of direct feedback scheduling scheme, since it determines the priorities of the control loops *directly* according to their current performance. Of course, this method is not actually a scheduling algorithm from the viewpoint of real-time scheduling theory. This is because we do not realize any real-time scheduler for operating systems. While the sensors' priorities are modified periodically, the network is still scheduled based on the fixed-priority scheduling algorithm at runtime. On the other hand, since new low-level schedulers are not needed for our approach, it is easy to implement in practice.

### 4.4. Design Considerations

Fig. 2 gives the pseudo code for the integrated feedback scheduling algorithm. In the following, the design of some critical parameters will be discussed.

Since the feedback scheduler is time triggered, an appropriate invocation interval $T_{FS}$ needs to be determined. It is intuitive that as $T_{FS}$ becomes smaller the feedback scheduler will become more sensitive to variations in available resource, which benefits the improvement of feedback scheduling performance. However, decreases in $T_{FS}$ will add computing and communication overheads. On the other hand, to achieve accurate measurements of deadline miss ratio, $T_{FS}$ should not be too small. In practice, when choosing $T_{FS}$, one often has to take into account a set of characteristics of the system, e.g., magnitudes of the sampling periods of the control loops and (estimated) frequency of workload variations, and make a trade-off between feedback scheduling overhead and sensitivity.

```
//ρ: Deadline miss ratio
//e: Control error (in control loop)
//h: Sampling period
//p: Loop/sensor priority
Integrated Feedback Scheduling {
    Input: ρ, {e}
    Period Adjustment {
        //Calculate new total utilization U
        Compute ERR using (3);
        Compute U using (2);
        //Reassign sampling periods
        FOR each control loop
            Compute performance index J using (5);
        END
        FOR each control loop
            Compute new sampling period h using (6) or (7);
        END
    }
    Priority Modification {
        FOR each control loop
            Compute J′ using (8) ;
        END
        Sort control loops with decreasing J′ values;
        Determine new priorities for sensors according to RULE 1-5;
    }
    Output: {h,p}
}
```

**Fig. 2.** Pseudo code for the integrated feedback scheduling

The setpoint for deadline miss ratio $\rho_r$ highly relates to the robustness of the control loops to packet losses. Recall that packet losses are only one subclass of deadline misses. Theoretically, one could first use related control theory (e.g. [6,9]) to obtain the maximum allowable packet loss ratio $\rho_{i,max}$, and then choose an appropriate $\rho_r$ in the range of (0, min$\{\rho_{i,max}\}$). Generally, it is not desirable to choose a $\rho_r$ value too close to min$\{\rho_{i,max}\}$. To maintain system stability, it is necessary to preserve an enough margin for dynamic variations of deadline miss ratio. However, no guarantee of system stability can be made explicitly even if a $\rho_r$ smaller than min$\{\rho_{i,max}\}$ is selected. Fortunately, to minimize the impact of deadline misses in steady states on QoC, most often we will specify a relatively small $\rho_r$, which makes it easy to provide stability guarantees.

Appropriate control parameters $K_P$ and $K_I$ must be chosen in (2). Ideally, if the relationship between deadline miss ratio and utilization can be described analytically, well-established feedback controller design methods such as pole placement could then be used to obtain these two parameters. However, due to the complexity of network communications, the relationship between the deadline miss ratio and utilization is unable to formulate analytically in most circumstances, particularly when NCSs operate in dynamic environments with uncertainty. Therefore, $K_P$ and $K_I$ are tuned by simulations.

In the module of priority modification, the value of the switch threshold $\delta$ affects the feedback scheduling performance to some degree. If it is too large, then the advantage of priority modification will be weakened undesirably. If it is too small, however, frequent switches of priorities cannot be avoided effectively. When choosing $\delta$, the runtime control performance of the system should be taken into account. It could often be determined based on the magnitudes of $J'_i$ values of the control loops.

## 5. Performance Evaluation

In this section we evaluate the performance of the proposed IFS scheme via simulation experiments. Suppose that the control network is of CAN type with a data rate of 25Kbps. Although the data rate of CAN buses can be much higher, it could be regarded that all other bandwidth resources have been allocated to other communication entities. For simplicity, all physical processes in the control loops are assumed to be DC motors. State feedback controllers are designed using the pole placement method. The relevant process model and desired closed-loop poles are given below.

DC motor: $\dot{x} = \begin{bmatrix} -1 & 0 \\ 1 & 0 \end{bmatrix} x + \begin{bmatrix} 1 \\ 0 \end{bmatrix} u, \quad y = \begin{bmatrix} 0 & 1 \end{bmatrix} x$

Desired closed-loop poles (on $z$ plane): $0.8 \pm 0.3i$

The size of sample data packets in all control loops is 10 bytes. The corresponding data packet transmission time is $10 \times 8/25 = 3.2$ms. Weighting coefficients $w_i = 1$ for all control loops. Some parameters of the feedback scheduler are listed in Table 1. Various scenarios with both light load and heavy load are simulated, and the results are compared with those from the traditional NCS design method (denoted Non-FS) that employs fixed sampling periods and the RM scheduling algorithm.

**Table 1.** Simulation parameters for integrated feedback scheduling

| Variable | $T_{FS}$ | $\rho_r$ | $K_P$ | $K_I$ | $h_{max}$ | $\varepsilon$ | $\delta$ |
|---|---|---|---|---|---|---|---|
| Value | 500ms | 5% | 0.3 | 0.8 | 20ms | 0.2 | 0.2 |

## 5.1. Scenario I: Underloaded Conditions

This scenario simulates the NCS with light network load. There are only two control loops in the system. Set $h_1 = 10$ms and $h_2 = 12$ms. The initial requested network utilization is $3.2/10 + 3.2/12 = 0.587$. It can be concluded from Theorem 1 that the system is schedulable. By default, loop 1 holds the high priority. At runtime the inputs to the control loops are square waves with periods of 4s and 2s respectively.

To quantify the QoC, Fig. 3 gives the integral of absolute error (IAE) for each of the control loops. Compared to the Non-FS case, IFS leads to 14.1% improvement in the control cost for loop 1, and 50.5% for loop 2. The total control cost improvement is 40.9%, indicating that the QoC under IFS is much higher than that under Non-FS.

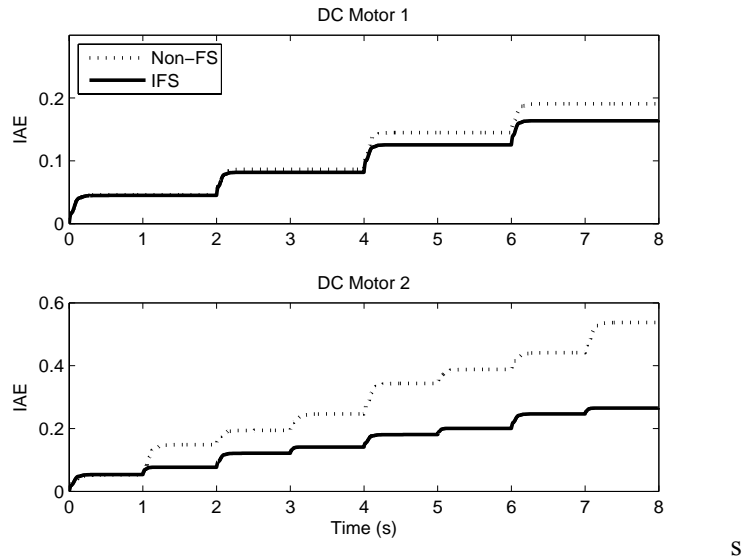

s

**Fig. 3.** Integral of absolute error for each control loop in Scenario I

As shown in Fig. 4, both the sampling periods and the priorities of control loops remain constant at runtime when the traditional design method is used. As a result, the allocation of network resource is fixed. Note that in all figures for priorities a smaller value represents a higher priority level.

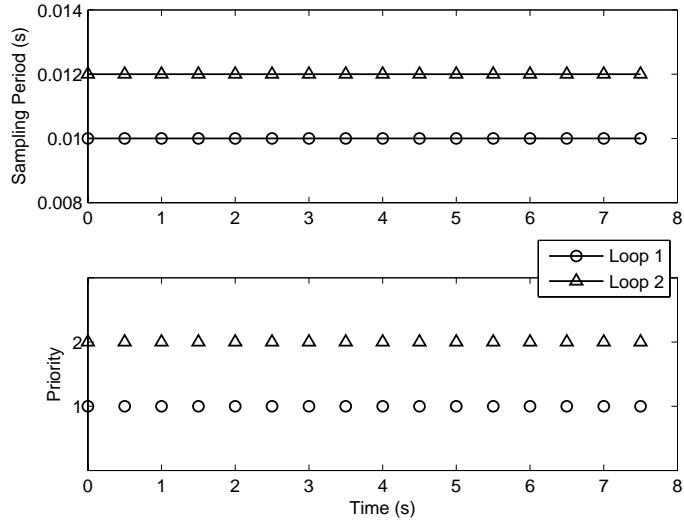

**Fig. 4.** Sampling periods and priorities under Non-FS in Scenario I

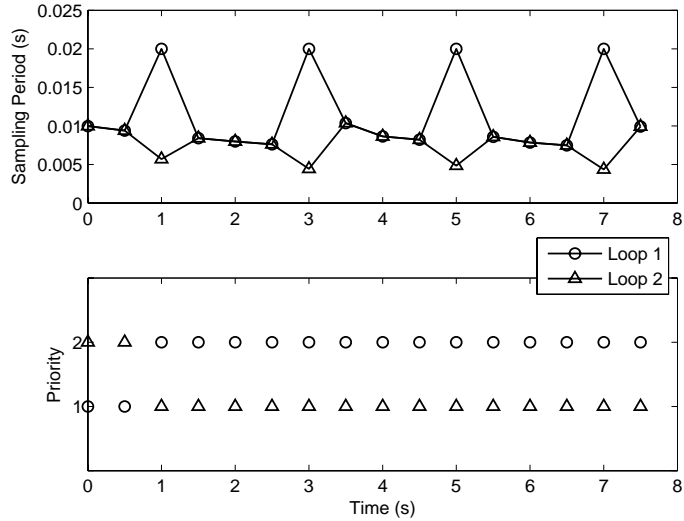

**Fig. 5.** Sampling periods and priorities under IFS in Scenario I

As can be seen from Fig. 5, under IFS the sampling periods and the priorities change over time. If two control loops have comparable performance, both sampling periods will be shortened. For instance, at time t = 6.5s, both control loops are in steady states, and consequently their sampling periods are 7.5 ms, which is smaller than both of their respective initial values (i.e. 10ms and 12ms). If the control performance of two loops is rather different, the feedback scheduler will assign a relatively smaller sampling period and a higher priority to the control loop with worse performance. For instance, at time t = 7s, loop 2 is experiencing a transient process while loop 1 is in a steady state. In order to bring loop 2 back to its steady state as soon as possible, the system assigns the loop a relatively small sampling period of 4.3ms and the high priority of value 1. At the same time, the sampling period of loop 1 is enlarged to 20ms so that the system schedulability is not violated. As a result, our IFS scheme results in improvement of the overall QoC.

When the traditional design method is employed, the total requested network utilization of control loops remains at a relatively low level of 58.7% at runtime, see Fig. 6. On the contrary, the requested network utilization under IFS increases gradually to around 80%, much higher than 58.7% under Non-FS.

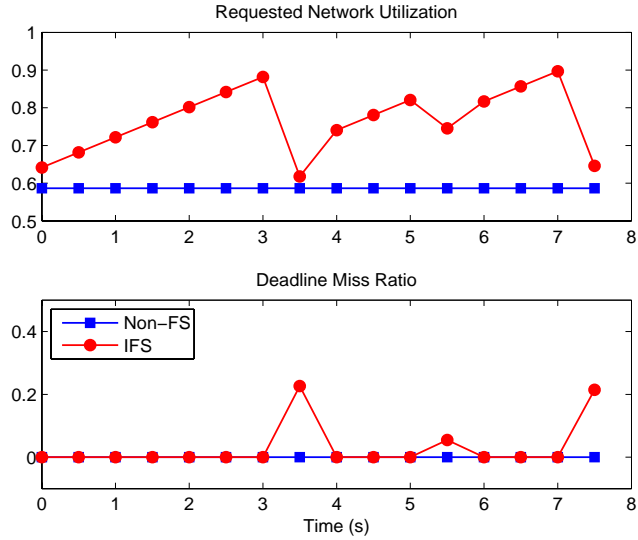

**Fig. 6.** Requested network utilization and deadline miss ratio in Scenario I

As also shown in Fig. 6, although deadline misses occur when the integrated feedback scheduler is used, the deadline miss ratio remains considerably small (even to be 0) most of the time, without jeopardizing the overall control performance.

### 5.2. Scenario II: Overloaded Conditions

This scenario simulates an overloaded NCS with four control loops, in which $h_1 = h_2 = 10$ms and $h_3 = h_4 = 12$ms. This scenario can be viewed as a consequence of adding two extra control loops onto the system considered in Scenario I for the purpose of system update. In this context the initial requested network utilization is $2 \times \left( \frac{3.2}{10} + \frac{3.2}{12} \right) \times 100\% = 117.4\%$, indicating that the system is overloaded and hence is not schedulable. Therefore, with traditional open-loop scheduling methods, deadline misses cannot be avoided. By default, the priorities are set as: $p_1 > p_2 > p_3 > p_4$. The inputs to loops 1 and 2 are square waves with a period of 4s, and loops 3 and 4 have the same square wave inputs with a period of 2s.

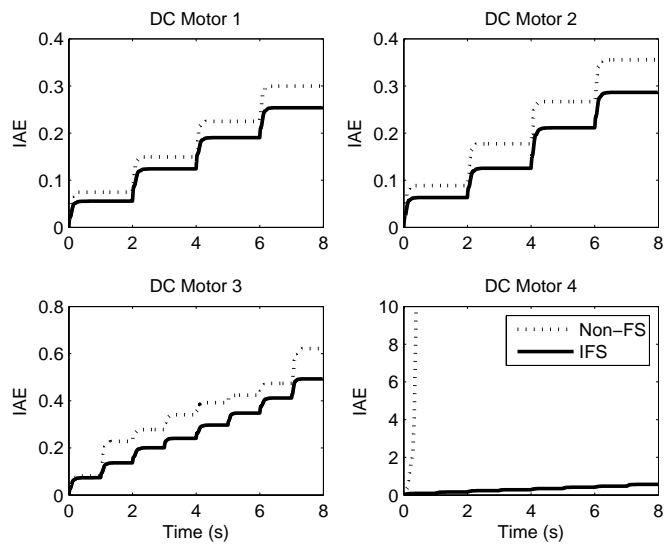

**Fig. 7.** Integral of absolute error for each control loop in Scenario II

Fig. 7 shows the IAE value for each control loop. With traditional open-loop scheduling, loop 4 finally becomes unstable, though the other three loops achieve satisfactory performance. Compared with the traditional design method, IFS yields improved QoC for all control loops. Quantitative evaluation in terms of IAE shows that the first three loops have the performance improvement of 15.2%, 19.3%, and 20.6%, respectively, and loop 4 is stabilized with the control performance comparable to that of loop 3.

We further analyze why improved results can be achieved. Under overloaded conditions, the integrated feedback scheduler makes the unschedulable system schedulable through adapting the sampling periods of the control loops. As can be seen from Fig. 8, if the performance of the control loops is comparable, the feedback scheduler will decrease the total requested network utilization through properly enlarging sampling periods, thus decreasing the deadline miss ratio of the system (see Fig. 10).

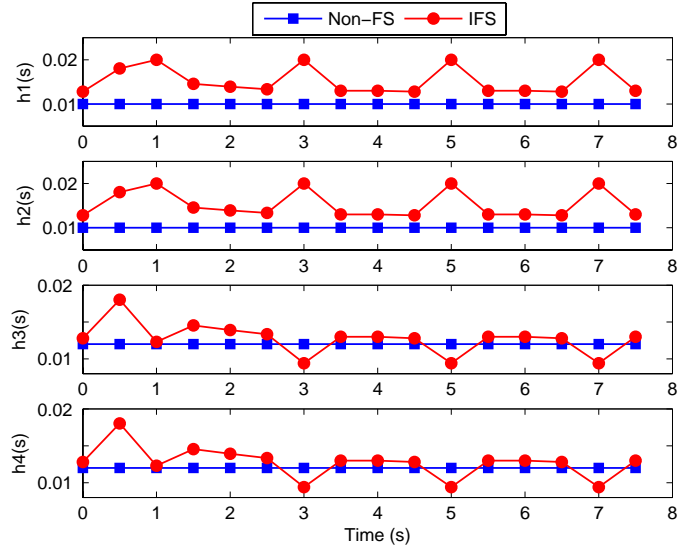

**Fig. 8.** Sampling periods of control loops in Scenario II

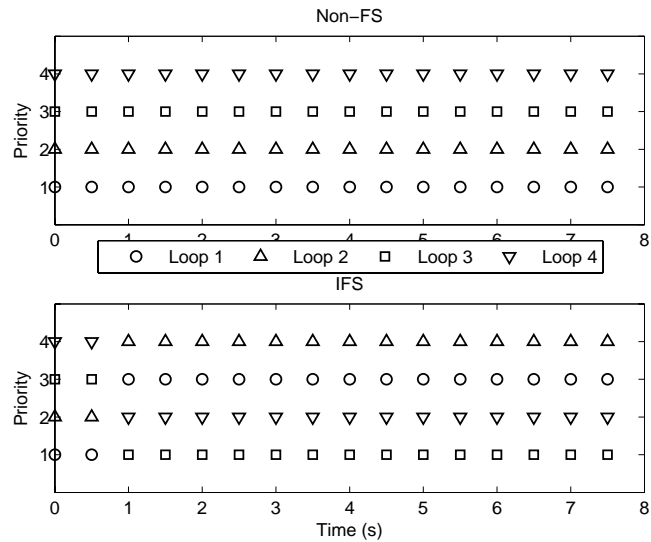

**Fig. 9.** Priorities of control loops in Scenario II

If the performance of the control loops is quite different, the feedback scheduler will assign relatively small sampling periods to the control loops with worse performance, and relatively large ones to the control loops with better performance. Besides sampling periods, the priorities of the control loops also vary under IFS, as shown in Fig. 9. In contrast, both sampling periods and priorities of all control loops are fixed with Non-FS.

Fig. 10 depicts the total requested network utilization and the deadline miss ratio under different schemes. Under Non-FS, the network workload remains to be 117.4% all the time, and the deadline miss ratio is around 25.5%. In the case of IFS, the requested network utilization keeps below (and close to) 100%, and the deadline miss ratio finally becomes very low (close to 0), both approaching steady states after a transient process.

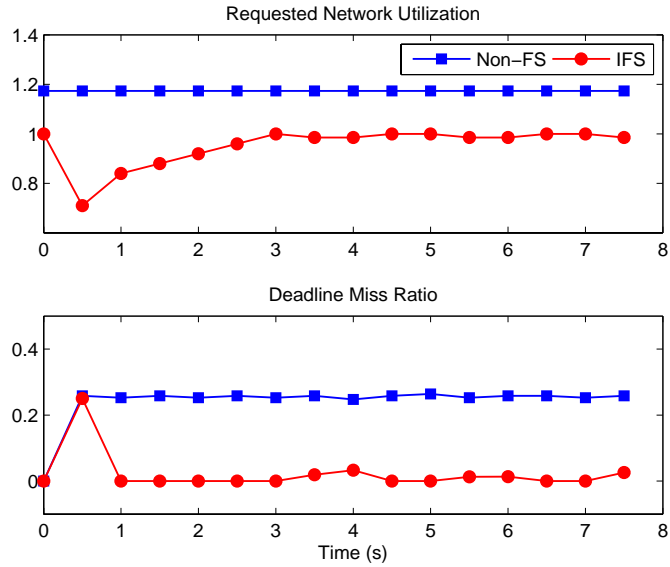

**Fig. 10.** Requested network utilization and deadline miss ratio in Scenario II

As mentioned above, to reduce the deadline miss ratio, IFS enlarges the average sampling period of each loop to some degree. Although in theory larger sampling periods will yield worse control performance, the total control cost of the system decreases because the deadline miss ratio is significantly reduced. Like in Scenario I, thanks to the dynamic adjustment of both sampling periods and priorities, which enables each control loop to obtain as much bandwidth as possible when it needs most, the performance of all control loops is improved.

## 6. Conclusion

By exploiting the emerging methodology of codesign of feedback control and network scheduling, an integrated feedback scheduling scheme has been proposed in this paper. It combines a cascaded feedback scheduling algorithm for sampling period adjustment and a direct feedback scheduling algorithm for priority modification. It optimizes the use of available network resources through dynamic adaptation of both sampling periods and priorities, thus enabling flexible QoC management in NCSs. With the integrated feedback scheduling, the deadline miss ratio of NCSs can be controlled effectively. The available network resources will be fully used even in underloaded conditions, while graceful degradation of QoC can be achieved in overloaded conditions. Simulation results have shown that the proposed scheme is able to effectively tackle the problem of bandwidth limitation and workload variations, thus providing a new approach to NCS design and implementation in dynamic environments. However, it is still non-trivial to examine the runtime overhead associated with integrated feedback scheduling. Advanced control techniques will be employed in the cascaded feedback scheduling algorithm in our future work.

## Acknowledgement

This work is supported in part by China Postdoctoral Science Foundation under Grant No. 20070420232 and Australian Research Council (ARC) under the Discovery Projects Grant No. DP0559111.